# Vortex rings drive entrainment and cooling in flow induced by a spark discharge

Bhavini Singh[1], Lalit K. Rajendran[1], Jiacheng Zhang[2], Pavlos P. Vlachos[2] and Sally P.M. Bane[1]

**Abstract**

Spark plasma discharges induce vortex rings and a hot gas kernel. We develop a model to describe the late stage of the spark induced flow and the role of the vortex rings in the entrainment of cold ambient gas and the cooling of the hot gas kernel. The model is tested in a plasma-induced flow, using density and velocity measurements obtained from simultaneous stereoscopic particle image velocimetry (S-PIV) and background oriented schlieren (BOS). We show that the spatial distribution of the hot kernel follows the motion of the vortex rings, whose radial expansion increases with the electrical energy deposited during the spark discharge. The vortex ring cooling model establishes that entrainment in the convective cooling regime is induced by the vortex rings and governs the cooling of the hot gas kernel, and the rate of cooling increases with the electrical energy deposited during the spark discharge.

## I. INTRODUCTION

There is a significant interest in understanding the dynamics of the flow induced by spark discharges. However, the induced flow field is complex, transient, and requires diagnostics with high spatiotemporal resolution. As a result, prior work has been limited to qualitative experiments [1]–[4] or computational models focused on the very early (<100 µs) stages of the induced flow field [5]–[8]. These studies have shown that the spark creates a hot gas kernel between the electrodes that cools over time. Some computations have shown the presence of vortices near the electrodes [4], [7], and in one of these studies, the authors postulate that these vortices are responsible for the shape of the hot gas kernel [4]. Recently, we performed measurements using stereoscopic particle image velocimetry (S-PIV) and background oriented schlieren (BOS) to measure the velocity and density fields, respectively, during the kernel cooling process [9], [10]. The velocity measurements showed the presence of vortex rings near each electrode and significant entrainment of ambient gas into the electrode gap [9], [11]. The density measurements showed that the cooling of the hot gas kernel occurs in two distinct regimes: an initially fast, convective cooling regime, followed by a slower cooling regime, with the fast regime contributing approximately 50% of the overall cooling in less than 1 ms after the spark discharge. We also developed a reduced-order model to describe the cooling process and showed that the entrainment of cold ambient gas

---

[1] School of Aeronautics and Astronautics, Purdue University, West Lafayette, IN 47907, USA
[2] School of Mechanical Engineering, Purdue University, West Lafayette, IN 47907, USA

into the electrode gap controls the rate of cooling in the fast regime [9]. However, the driving mechanism behind the cold gas entrainment remains unresolved.

In this work, we use theory and experiment to demonstrate that the induced vortex rings drive the entrainment and thus control the rate of cooling of the hot gas kernel. We develop a model for the entrainment using inviscid vortex ring theory that directly relates the geometric and kinematic properties of the vortex rings to the cooling. To test the model predictions, we perform simultaneous measurements of velocity and density in a flow field induced by a spark discharge using time-resolved S-PIV and BOS. Simultaneous measurements are critical because the spark discharge is a chaotic process, and thus each realization of the flow field is different.

The high spatiotemporal resolution measurements capture the spatial organization of the vortex rings with respect to the hot gas kernel. We reveal the role that induced vorticity plays in heat transfer and the dynamics of flows induced by sparks and discover that spark discharges produce vortex-driven mixing flow. This could have broad implications in a variety of scientific, engineering and technological applications by enabling the alteration of temperature, chemical species distribution (combustion), and momentum transport (flow control).

## II. VORTEX RING COOLING MODEL

The vortex ring cooling model presented herein expands upon our prior work, where a model relating entrainment to the cooling of the hot gas kernel was developed by considering the rate of change of enthalpy of the kernel [9]. We assumed low Mach number, inviscid flow, negligible body forces, negligible heat transfer due to conduction and radiation, as well as a calorically perfect gas, to calculate the *cooling ratio* parameter as:

$$\left(\frac{\overline{\rho_f} - \overline{\rho_i}}{\rho_\infty - \overline{\rho_i}}\right) = 1 - exp\left(-\int_{t_i}^{t_f} \frac{\dot{V}_{in}}{V_{cv}} dt\right) \quad (1)$$

where $\bar{\rho}$ is the mean kernel density, and $t$ is time, with the subscripts $i$ and $f$ representing the initial and final values in the convective cooling regime. The ambient density is given by $\rho_\infty$, while $\dot{V}_{in}$ is the volumetric entrainment, and $V_{cv}$ is the volume of fluid enclosed by the control volume. The cooling ratio represents the extent of completion of the cooling process, with a value of 1 indicating the fluid has cooled to the ambient temperature. The present work aims to relate the entrainment ($\dot{V}_{in}$) to the properties of the vortex rings, subject to the hypothesis that the pair of vortex rings drive the entrainment and therefore govern the cooling in the convective regime.

### A. Entrainment model

We approximate the hot gas kernel as a moving and deforming cylindrical control volume, defined by the centroids of the two vortex rings at the top and bottom and centered at the electrode axis, $r = 0$ (Figure 1 (a)). The top and bottom vortex rings bounding the kernel are located at $Z_t$ and $Z_b$ and have ring radii $R_t$, $R_b$, and core radii $a_t$ and $a_b$, respectively. Entrainment into the control volume is calculated across the top and bottom boundaries. In the experimental measurements, the top and bottom vortex rings do not have the same dimensions, so the radius of the cylinder is taken as the mean of the top and bottom ring radii.

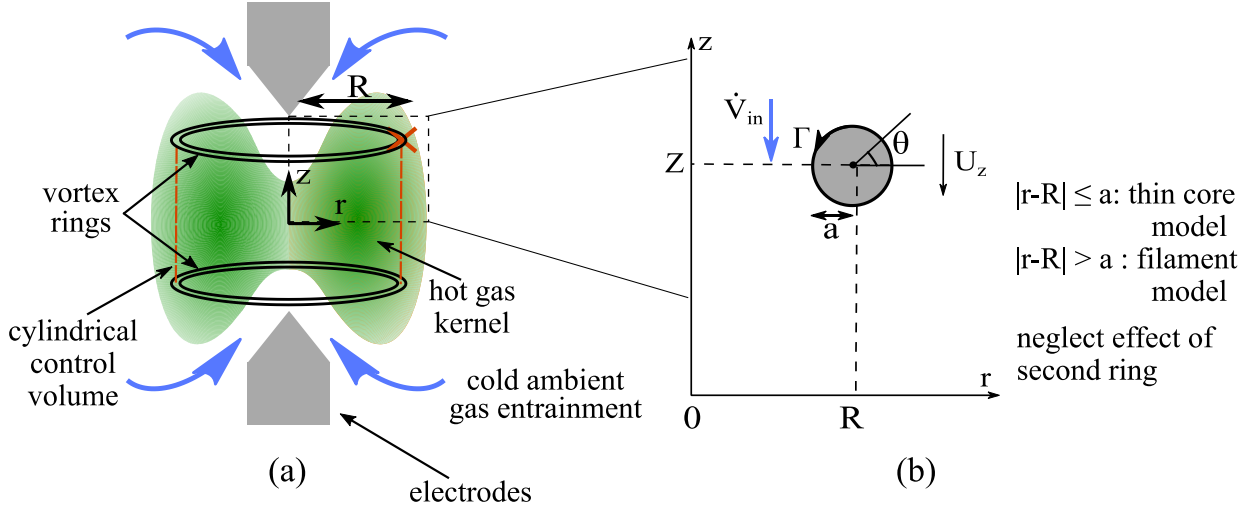

**Figure 1: (a) Schematic of the vortex ring cooling model showing entrainment of cold ambient gas into the hot gas kernel, hypothesized to be due to a pair of thin core vortex rings. (b) Schematic showing a thin core ring with ring radius $R$, core radius $a$, circulation $\Gamma$ and $z$-component of ring velocity $U_z$. Entrainment ($\dot{V}_{in}$) is calculated using the thin core model inside the core and filament model outside. The effect of the second ring on entrainment across the boundary of the first ring is negligible, and vice versa.**

We calculate the entrainment using a coordinate system attached to the vortex ring, as follows:

$$\dot{V}_{in} = 2\pi \left( \int_0^R u_z r \, dr - \int_0^R U_z r \, dr \right) \tag{2}$$

where $R$ is the mean ring radii of the top and bottom vortex rings $\left( R = \frac{1}{2}(R_t + R_b) \right)$, and $U_z$ is the $z$-component of the ring velocity, which is calculated separately for the top and bottom rings. The total entrainment is the sum of the entrainment across both the top and bottom boundaries of the control volume.

We define a function $f(r,z)$, the entrainment function in cylindrical coordinates, such that:

$$u_r = -\frac{1}{r}\frac{\partial f}{\partial z} \quad \text{and} \quad u_z = \frac{1}{r}\frac{\partial f}{\partial r} \tag{3}$$

where, $u_r$ and $u_z$ are the radial and axial velocities, respectively. Substituting the definition of the entrainment function (Equation 3) into Equation 2 gives:

$$\dot{V}_{in} = 2\pi \left( [f(R,Z) - f(0,Z)] - \frac{1}{2} U_z r^2 \Big|_0^R \right) \quad (4)$$

For incompressible flow, this entrainment function is equivalent to the stream function. The low Mach number assumption implies that changes in density are only due to changes in temperature and that the flow is acoustically incompressible (as the cooling process occurs long after the compression wave has departed the field of view). Buoyancy effects are also negligible. Under these assumptions, we can employ vortex ring models derived for incompressible flows to calculate entrainment for our flow field, assuming one-way coupling, namely that variations in density do not affect the velocity field.

### B. Thin core vortex ring model

The entrainment is calculated as the difference of the values of the entrainment function between the ring centroid ($r = R$) and the axis of symmetry ($r = 0$) (Equation 4). Using ring-fixed coordinates, the motion of the ring is accounted for by subtracting the last term in Equation 4. The total entrainment into the control volume is the sum of the entrainment due to both vortex rings. Several approaches have been proposed to model the entrainment induced by vortex rings [12]–[17], and they depend on the model used for the vorticity distribution inside the rings. In this work, we consider inviscid, thin-cored vortex rings of uniform vorticity inside the core ($\omega(r,z) = \omega_0 r/R$) where it is assumed that the cores do not overlap, and the core radius is much less than the ring radius. We use two models to calculate the entrainment function: 1) a filament model outside the core of the ring and 2) a thin-core model for points that lie inside the ring core, as shown in Figure 1(b). Since the equation for entrainment (Equation 4) employs the value of the function at two points, we use the filament model at the axis of symmetry ($r = 0$), and the thin-core model at the ring centroid ($r = R$). Finally, the entrainment due to one vortex ring at the location of the other ring was found to be negligible (on average, 10% of the entrainment due to the second ring on itself) and was therefore omitted in the total entrainment calculation. In summary, the entrainment across the top and bottom faces of the control volume is calculated based only on ring properties.

The entrainment function, according to the filament model, is given by [13], [16], [18]:

$$f(r,z) = \frac{\Gamma}{2\pi} \sqrt{Rr} \left[ \left( \frac{2}{\sqrt{m}} - \sqrt{m} \right) K(m) - \frac{2}{\sqrt{m}} E(m) \right] \quad (5)$$

where $\Gamma$ is the circulation of the vortex ring, $m = \frac{4rR}{L^2}$, $L^2 = (z - Z)^2 + (r + R)^2$ and $K$, $E$ are elliptic integrals of the first and second kind. The filament model assumes inviscid, infinitesimally thin vortex rings and is valid at distances far from the centroid of the vortex ring.

The entrainment function inside the core of the vortex ring is calculated using the thin-core model, according to Saffman [13]. This model uses the asymptotic properties of the elliptic integrals as $m \to 1$ ($r/R \to 1$) and assumes uniform vorticity inside the core to obtain:

$$f(r,z) = \frac{\Gamma R}{2\pi}\left[\log\left(\frac{8R}{a}\right) - \frac{3}{2} - \frac{d^2}{2a^2} + \frac{d}{2R}\cos\theta\left[\log\frac{8R}{a} + 1 - \frac{5d^2}{4a^2}\right]\right] \quad (6)$$

where $\theta$ is defined as shown in Figure 1 and $d$ is the distance from the centroid of the ring core and is given by $d = \sqrt{(r-R)^2 + (z-Z)^2}$.

The ring velocity $U_z$ in Equation 4 is taken to be the self-induced ring velocity [19] because the effect of the second ring is negligible:

$$U_z = \frac{\Gamma}{4\pi R}\left[\log\frac{8R}{a} - \frac{1}{4}\right] \quad (7)$$

The model parameters, $\Gamma$, $R$, and $a$ are all calculated from the velocity field induced by the spark discharge measured using S-PIV experiments.

### III. EXPERIMENTAL METHODS AND TECHNIQUES

The cooling effect in a single spark discharge was investigated using measurements of the induced velocity field and density field. Simultaneous time-resolved S-PIV and BOS measurements were performed for a range of energy values from 2.2 to 5.1 mJ. Each experimental run was spaced out at least 30 seconds apart to eliminate residual flow effects from one spark event to the next.

#### A. Plasma generation

A single nanosecond spark discharge was generated between two electrodes using a high voltage pulse generator from Eagle Harbor Technologies. The electrodes were machined out of ceriated-tungsten and had cone-shaped tips. Details of the plasma generation set-up are given in prior work [9]. The energy deposited in the electrode gap was controlled by limiting the current through the discharge by adding resistances in series with the spark, as shown in Figure 2 where the label "R" represents the resistances. Total resistance values of 0, 100, 200, 400, 600, and 1000 were used in these experiments, where a resistance value of 0 means that no resistance was added in the electrical configuration. For each resistance level, multiple tests were conducted, resulting in a range of energy values from 2.2 mJ to 5.1 mJ.

## B. Simultaneous velocity and density measurements

### 1. Experimental set-up

A schematic of the simultaneous S-PIV and BOS system is shown in Figure 2. The time-resolved S-PIV system consisted of a QuasiModo quasi-continuous burst-mode (QCBM) Nd:YAG laser [20], two high-speed Photron SA-Z cameras, and an acrylic test section containing the electrodes. The laser was operated at 50 kHz, and laser sheet optics produced an approximately 1 mm thin waist in the region of interest where the spark discharge was generated. The cameras with an included angle of approximately $50^0$ were operated with Nikon AF FX Micro-Nikorr 200 mm lenses and recorded particle images at a sampling rate of 50 kHz. The resolution for the two cameras was approximately 600 x 640 pixels, corresponding to an approximate field of view of 6 x 6.5 mm. Alumina particles were injected into the test enclosure with diameters of about 0.3 µm and the estimated Stokes number of approximately 0.002. The BOS system consisted of a Phantom v2640 camera with a Nikon Nikorr 105 mm lens, f# 11, and a 2x teleconverter, which recorded at a 20 kHz sampling rate, with a resolution of 1024 x 704 pixels, corresponding to an approximate field of view of 10 x 7 mm. The dot pattern contained dots 42 µm in diameter with an equal edge to edge spacing between the dots of 42 µm, and was back-illuminated using a 150 W xenon arc lamp (Newport 66907). The distance from the center of the electrodes to the BOS target ($Z_D$) was approximately 32 mm, and the distance from the center of the electrodes to the camera ($Z_A$) was approximately 165 mm. A delay generator was used to synchronize and trigger the laser, the S-PIV, and BOS cameras and the high voltage pulser.

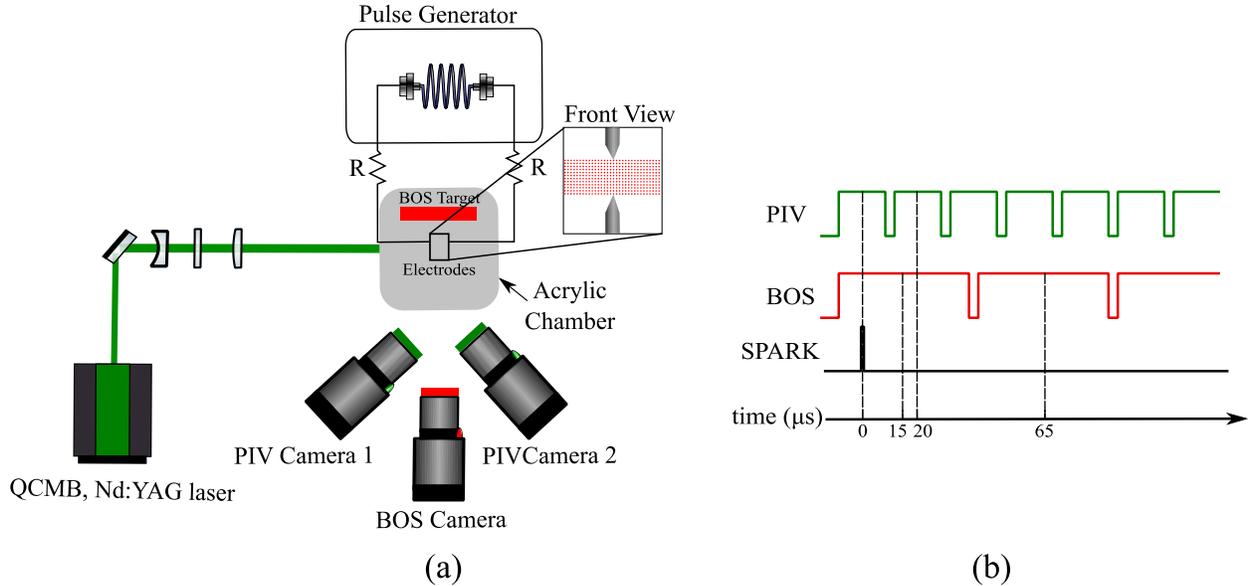

**Figure 2:** (a) Schematic of the experimental set-up for simultaneous S-PIV and BOS measurements of the plasma-induced flow field. (b) Timing diagram used to synchronize the simultaneous measurements.

## 2. S-PIV image processing and uncertainty quantification

The recorded S-PIV particle images were processed using PRANA (PIV, Research, and ANAlysis) software [21]. A total of four passes were used, and a 50% Gaussian window was applied to the original window size [22], resulting in window resolutions of 64x64 pixels in the first pass to 32x32 pixels in the last pass, with 50 % window overlap in all passes. The projected velocity fields calculated in this manner were then combined with the camera angle information obtained from calibration to yield the three components of velocity in the measurement plane [23]. Proper orthogonal decomposition with the entropy line fit (ELF) method [24] was used to denoise the PIV velocity fields. The spatial resolution for velocity calculations was 0.32 mm, and each snapshot contains 38 x 40 vectors. Further details on the calibration procedure and processing can be found in prior work [9].

Experimental uncertainties in the S-PIV measurements were propagated according to the procedure outlined by Bhattacharya *et al.* [25]. Uncertainties in the planar velocity fields were calculated using the moment of correlations (MC) [26] and then propagated through the stereo reconstruction to calculate uncertainty in the reconstructed velocity field. Maximum uncertainties in the three components of velocity (u, v and w) were approximately 0.1 m/s, 0.1 m/s and 0.3 m/s, corresponding to 2.5%, 2.5% and 7.5% of the maximum velocity (4 m/s), respectively.

## 3. BOS image processing and uncertainty quantification

The distortion of the dot pattern is estimated by cross-correlating an image taken with the flow to a reference image taken without the flow. The images were processed using the same approach and software (PRANA) as for the S-PIV measurements. A total of four passes were used with window resolutions of 64x64 pixels in the first pass to 32x32 pixels in the last pass, with a 50 % window overlap in all passes. The spatial resolution of the final pass was 0.31 mm, and each snapshot contained 62 x 24 vectors. The projected density field $\rho_p(x,y)$, was then calculated from the displacement field by 2D integration using an uncertainty based Weighted Least Squares (WLS) methodology, wherein the density gradient measurements are weighted based on the inverse of their uncertainty [27]. Finally, the actual density field $\rho(r,z)$ was calculated from the projected density field by Abel inversion under the assumption that the flow is axisymmetric. Further details on the processing and calculation of the density field from displacement values can be found in prior work [9].

Experimental uncertainties in the BOS measurements were propagated through the complete density reconstruction according to [28]. Uncertainties in the displacement field were calculated using MC [26], then propagated through the optical layout followed by the WLS solver to obtain uncertainty in the projected density field. The uncertainties were then propagated through the Abel inversion to obtain total uncertainty in the density field. The maximum uncertainties in the density field were approximately 0.005 kg/m$^3$, corresponding to 0.4% of ambient density (1.225 kg/m$^3$).

## C. Non-dimensionalization of results

All results were non-dimensionalized using the induced velocity ($u_{shock}$) and density ($\rho_{shock}$) behind the compression wave produced by the spark and the characteristic length ($d$) taken to be the electrode gap distance (5 mm) [9], [11]. The velocity and length scales were used to define a characteristic time scale, $\tau$, as $d/u_{shock}$. For the maximum energy value of 5.1 mJ, $u_{shock}$, $\rho_{shock}$ and $\tau$ are 61 m/s, 1.45 kg/m³, and 0.08 ms, respectively, and for the minimum energy value of 2.2 mJ, they are 41 m/s, 1.38 kg/m³, and 0.12 ms, respectively. Further details on the non-dimensionalization can be found in prior work [9].

## D. Entrainment function calculation from PIV velocity fields

The entrainment calculated from the vortex ring models was compared to a direct calculation of entrainment from the experiment according to Equation 4. The gradients of the entrainment function $f$ were calculated from the measured velocity field as:

$$\frac{\partial f}{\partial z} = -r u_r \text{ and } \frac{\partial f}{\partial r} = r u_z. \tag{8}$$

The field of $f$ was then obtained by solving the following equation as:

$$\begin{bmatrix} G_z \\ G_r \\ L_0 \end{bmatrix} f = \begin{bmatrix} -r u_r \\ r u_z \\ f_0 \end{bmatrix}, \tag{9}$$

where $G_z$ and $G_r$ are the discrete gradient operators (2D matrices) along the $z$-dimension and $r$-dimension, respectively, $f$ is the column vector containing the discrete entrainment function values, and $L_0$ is the labeling matrix consisting of zeros and ones for imposing the Dirichlet boundary conditions $f_0$. Equation 9 forms an overdetermined linear system and may not be consistent with possessing a solution. Thus Equation 9 was solved via optimization using least-squares as:

$$f = \underset{f}{\mathrm{argmin}}(\|Gf - s\|_2) = (G^T G)^{-1}(G^T s),$$
$$G = [G_z \quad G_r \quad L_0]^T$$
$$s = [-r u_r \quad r u_z \quad f_0]^T \tag{10}$$

where $G$ represents the augmented linear operator and $s$ is the augmented source term.

The variances $\sigma_s^2$ of the values in $s$ can be estimated by propagating the uncertainty of the velocity measurement through Equation 8 as:

$$\boldsymbol{\sigma_s} = [r\sigma_{u_r} \quad r\sigma_{u_z} \quad \sigma_{f_0}]^T \tag{11}$$

where $\sigma_{u_r}$ and $\sigma_{u_z}$ are the uncertainty of the velocity components $u_r$ and $u_z$, respectively, and $\sigma_{f_0}$ is the uncertainty of the Dirichlet values in $f_0$. With the assumption that the errors in the source term are uncorrelated, the covariance matrix $\Sigma_s$ of the augmented source term can be estimated as the diagonal matrix with the diagonal elements being $\boldsymbol{\sigma_s^2}$. Thus, the uncertainty of the calculated entrainment function can be obtained by propagating $\Sigma_s$ through Equation 10 as:

$$\Sigma_f = M\Sigma_s M^T,$$
$$with\ M = (G^T G)^{-1} G^T, \tag{12}$$

where $\Sigma_f$ is the covariance matrix of the entrainment function. The square root of the diagonal elements in $\Sigma_f$ corresponds to the uncertainty in the calculated entrainment function values.

### E. Post-processing metrics and calculation of parameters

The post-processing steps are summarized in Figure 3. Once the density and velocity fields were processed, the 2D grid was split along the electrode axis ($r = 0$), and the left side was flipped and combined with the right-hand side to create a spatially averaged axisymmetric field. The velocity field was then time averaged to match the temporal resolution of the density field. The coherent vorticity was calculated from the velocity field using a 4th order compact noise optimized Richardson extrapolation scheme [29], and the coherent structures were identified based on swirl strength using the $\lambda_{ci}$ method [30]. All swirl values that were at least 2% of the maximum swirl strength were taken as the vortex cores. The core radius was defined as the radius of a circle that occupied the same area as the vortex cores, and the centroid of the core was defined as the barycenter of the calculated area, weighted by $\lambda_{ci}$ values. These centroids were used to define the cylindrical control volume used in the calculation of both the mean density and entrainment from the BOS and S-PIV measurements, respectively. The mean density within the control volume was calculated, and the time history of the mean kernel density was used to determine the cooling regimes in the flow. The cooling analysis in this work is restricted to the fast, convective cooling regime, which ranges from $t = t_i$, the time of minimum density, to $t = t_f$, the changeover point, or the end of the convective cooling regime. The changeover point, $t_f$, is the time instant at which the running mean and slope of the kernel density change abruptly. The cooling rate in the convective cooling regime is the slope of the fitted line to the mean density in this regime.

The experimental entrainment function values from S-PIV data were interpolated at the ring centroid locations defining the control volume as well as at $r = 0$ and used to calculate the total, instantaneous entrainment for all tests. The tracked vortex ring centroids were differentiated in time to determine the experimental ring velocity $U_z$ in Equation 4. The vortex ring parameters, $R, a,$ and $\Gamma = \oint \vec{u} \cdot \vec{dl}$ were also calculated at each time step for each identified vortex ring and used to calculate the total instantaneous entrainment from the vortex ring model.

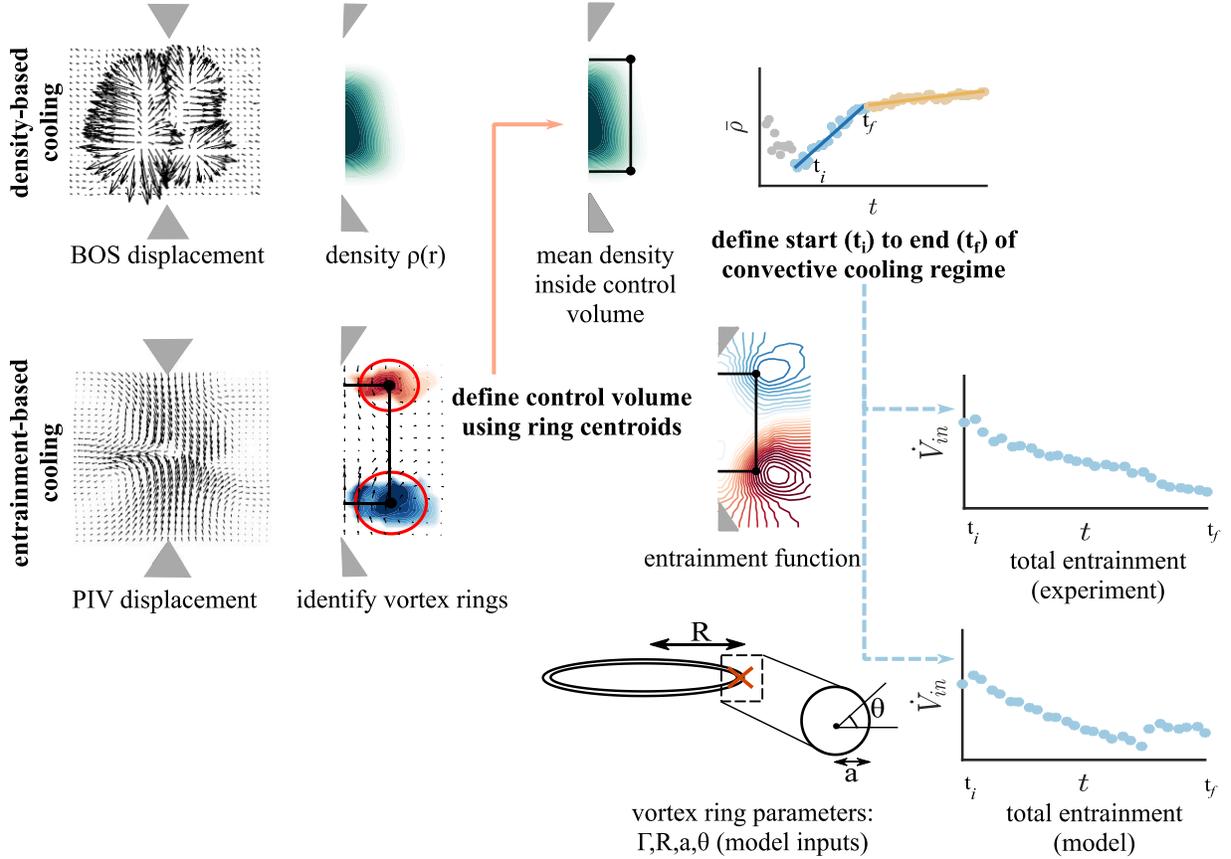

Figure 3: Post-processing steps for S-PIV and BOS measurements showing the identification of coherent vortex rings and density calculation. The centroids of the identified vortex rings are used to define the control volume used for entrainment calculations from experiment and model as well as mean kernel density calculation from BOS.

## IV. RESULTS AND DISCUSSION

### A. Spatial organization of the flow field and the effect of energy

The plasma-induced flow field for the highest (5.1 mJ) and lowest (2.2 mJ) energy cases are shown in Figure 4. The figure superimposes the hot gas kernel, shown in green, with the vortex rings and the experimentally calculated entrainment function.

In the initial stages of flow evolution ($t/\tau = 2$), the shock wave has departed the field of view, and a cylindrical region of hot gas is observed. A pair of vortex rings are also induced and located close to the periphery of the hot gas kernel. The pair of vortex rings then move towards each other due to self-induction, eventually colliding, as the hot gas kernel simultaneously compresses axially. At $t/\tau = 10$, the vortex ring centroids are located entirely within the hot gas kernel. The rings and hot gas kernel then begin to expand radially outward while the hot gas kernel continues to cool, its minimum density reaching 90% of ambient density by $t/\tau = 27$.

The minimum density in the kernel ($\rho/\rho_{shock}$), for the high and low energy cases, at $t/\tau = 2$ is 0.48 and 0.59, respectively. The circulation of the vortex ring in the high energy case is 3 times larger than the low energy case. The vortex rings and hot gas kernel expand, with their radial extent ($r/d$) increasing from 0.28 and 0.16 at $t/\tau = 2$ to 0.55 and 0.3 by $t/\tau = 27$ for the high and low energy cases, respectively, with a corresponding reduction in kernel height by a factor of 4 and 2 for the two different energy values.

These results underscore that there is a coupling between the dynamics of the vortex rings and the cooling of the hot gas kernel. We further demonstrate that the energy deposited in the gap affects the level of heating in the electrode gap and both the strength of the vortex rings and the extent of radial expansion of the rings and hot gas kernel. The higher energy case induces a hot gas kernel that expands out radially along with the vortex rings, to a radius of approximately half the electrode gap distance, forming a torus shape. Conversely, in the lower energy case, both the vortex rings and hot gas kernel remain concentrated in the center of the gap with much less radial expansion, approximately half of that in the higher energy case. Stronger vortex rings and lower density (more heating) are observed in the high energy case compared to the lower energy case. More heating in the high energy cases leads to more significant density gradients between the kernel and ambient, resulting in higher cooling rates of the hot gas kernel. Similarly, the mean entrainment values also increase with energy, because the increase in energy results in higher induced velocities within the electrode gap.

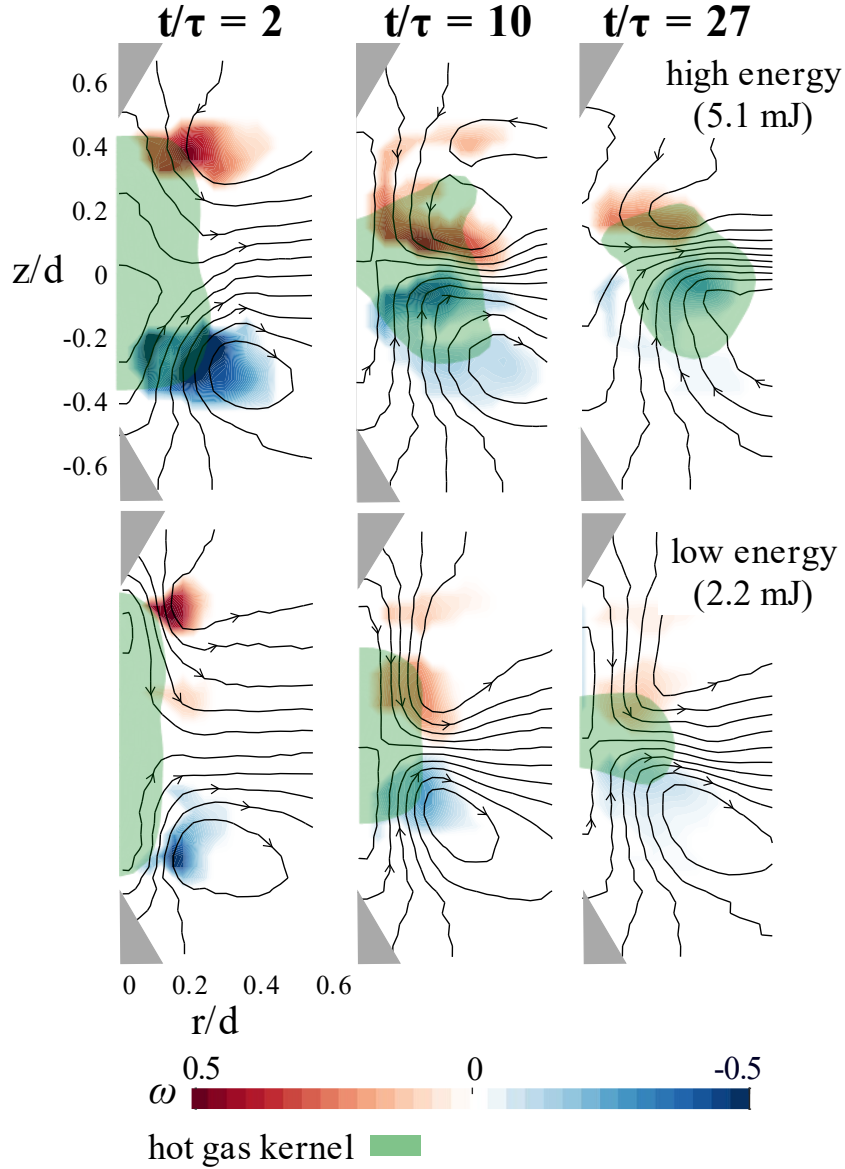

**Figure 4: Time evolution of the induced hot gas kernel and vortex rings for the highest energy case (5.1 mJ, top) and the lowest energy case (2.2 mJ, bottom). A portion of the pair of vortex rings is located inside a hot gas kernel for all tests, with the spatial distribution of the vortex rings controlling the shape of the hot gas kernel.**

## B. Application of vortex ring cooling model

The fast, convective cooling of the hot gas kernel is due to entrainment of cold ambient gas [9], but the driving mechanism behind this entrainment is not known. The simultaneous measurements in this study highlight the coupled behavior of the cooling of the hot gas kernel and the strength and dynamics of the vortex rings. These measurements also enable us to develop a model to test the hypothesis that the pair of vortex rings drive entrainment and control the cooling of the hot kernel. The model calculations were tested on two levels: comparison of (1) entrainment and (2) cooling ratio, calculated from the model to the values calculated directly from the experiment.

### 1. Entrainment

The instantaneous entrainment from the model and the S-PIV experiment calculated using Equation 4 is shown in Figure 5 (a). The entrainment from the model is calculated by substituting Equations 5-7 into Equation 4, and the entrainment from the experiment is calculated by substituting the entrainment function from the least-squares integration of the velocity field from the S-PIV measurements, described in Section III.D, into Equation 4. The mean difference of -1.2 $\times 10^{-5}$ between the two models, offset from the 1:1 line, shows that, on average, the model and experiment agree well. This mean difference between the model and experiment is within the mean uncertainty of the estimates (9 $\times 10^{-5}$ and 1.5 $\times 10^{-4}$, respectively). The entrainment values calculated from the model and experiment are statistically equivalent, supporting the hypothesis that the entrainment is due to the vortex rings.

### 2. Cooling

The results in Figure 5 (b) show the cooling ratio values calculated in 3 different ways: (1) The density-based method which is a direct calculation from experimental density measurements from BOS, (2) the entrainment-based method using the least-squares integration of the S-PIV measurements, and (3) the entrainment-based method using the vortex ring model, driven by data from the S-PIV measurements. The cooling ratios from the entrainment-based calculations from the vortex ring model and the least-squares integration of the S-PIV measurements are compared to those from the density-based calculations, from BOS measurements. There is good agreement between the entrainment-based cooling ratio calculations from the vortex ring model and the experiment with a mean difference of 0.003, which is within the mean uncertainty in the cooling ratio calculated from the experiment of 0.004 and the model of 0.005. Comparison of the cooling ratios from the entrainment-based calculations to density-based calculations, which are a direct measure of the cooling, shows that both the entrainment-based calculations underpredict the cooling ratio by 0.07 compared to the density-based results. This difference in cooling ratio of 0.07 is 14% of the mean cooling ratio estimated using the density-based method. Overall agreement between the cooling ratios from the experimental entrainment-based and density-based calculations confirm the results of our prior work that entrainment of cold ambient gas drives cooling of the hot gas kernel [9]. In addition, the agreement between the entrainment-based cooling ratio from the model and the density-based cooling ratio further supports the hypothesis that the vortex rings drive the cooling of the hot gas kernel. The cooling ratio values show that less cooling occurs at lower energy values. It was also observed (not shown) that the cooling occurs more rapidly with an increase in energy, with the cooling time for energy values less than 2.8 mJ observed to be about 1.6 times the duration at higher energy values. The cooling time at high energy values was approximately 1.2 ms and similar to previous work for the same energy range [9].

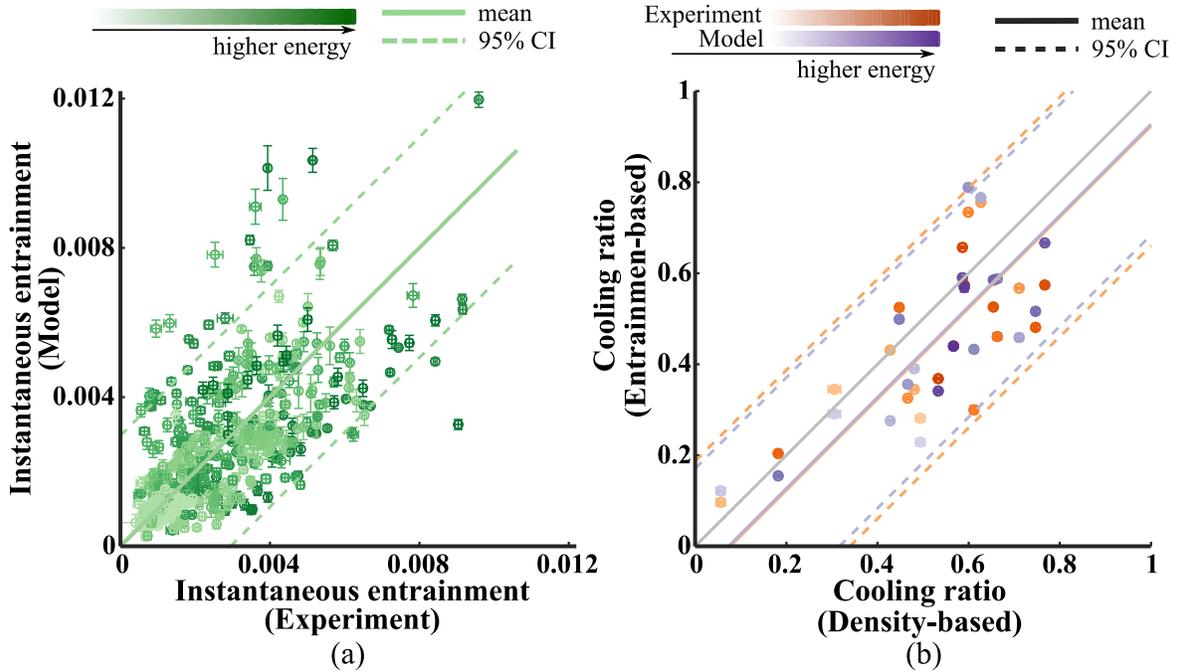

**Figure 5:** (a) Comparison of the total instantaneous entrainment calculated from the vortex ring model to experimental values calculated from least-squares integration of the S-PIV measurements and (b) comparison of cooling ratios from the entrainment-based calculations from the vortex ring model and the least-squares integration of the S-PIV measurements to the cooling ratios from the density-based calculations from BOS density data.

## CONCLUSIONS

The flow induced by a spark discharge consists of a pair of induced vortex rings and a hot gas kernel. In this work, we study this flow using simultaneous, high spatiotemporal-resolution, density, and velocity field measurements using BOS and S-PIV, respectively, to test the hypothesis that vortex rings drive entrainment and cooling in the induced flow field. To this end, we develop and test a vortex ring cooling model to describe the kernel dynamics using energy conservation and inviscid, axisymmetric, vortex ring models.

The vortex ring cooling model was tested by comparing entrainment and cooling calculated from the model to the experiment. On average, the entrainment calculated from the model and experiment agreed well, with the mean difference between the model and experiment being less than the experimental uncertainty. The cooling calculations of the model were evaluated against cooling calculated from PIV data and BOS data. The model once again agreed well with the PIV experiment, with the mean difference in cooling ratio values lying within experimental uncertainty. The model underpredicts the cooling ratio compared to BOS, with the mean difference being 14% of the mean BOS cooling ratio. The mismatch between the model and the BOS experiment may arise from the axisymmetric assumption made in developing the model. Though the electrode geometry suggests that the flow would be axisymmetric, the spark formed between the electrodes is rarely a perfect cylinder, and the induced shock wave and heating from the spark are not always axisymmetric. A complete understanding of the 3-dimensionality of the flow field will require volumetric measurements.

The results of this work discover that spark discharges induce vortex-driven mixing flows. The induced vortex rings govern the spatial distribution of the hot gas kernel by driving entrainment of cold ambient gas into the kernel, thereby cooling it. The energy deposited in the electrode gap affects the strength of the vortex rings, as well as the extent to which the rings and hot gas kernel expand radially. Stronger vortex rings are induced at higher energy values, resulting in more entrainment and faster cooling of the hot gas kernel. Therefore, the energy deposited in the electrode gap affects the strength and dynamics of the vortex rings, and hence the duration and extent of cooling/mixing, allowing one to control the late stages of the induced flow.

This detailed understanding of the role of the vortex rings in the mixing and cooling process of the hot gas kernel has wide-ranging implications for momentum transport and passive scalar mixing in plasma-based flow and combustion control techniques. The model can help guide further research on a variety of plasma discharges where the presence of vortex rings has been established [31]–[33].

## ACKNOWLEDMENTS

This material is based upon work supported by the US Department of Energy, Office of Science, Office of Fusion Energy Sciences under Award No. DE-SC0018156.

# References

[1] D. A. Xu, D. A. Lacoste, D. L. Rusterholtz, P. Q. Elias, G. D. Stancu, and C. O. Laux, "Experimental study of the hydrodynamic expansion following a nanosecond repetitively pulsed discharge in air," *Appl. Phys. Lett.*, vol. 99, no. 12, pp. 2009–2012, 2011.

[2] A. Borghese, A. D'Alessio, M. Diana, and C. Venitozzi, "Development of hot nitrogen kernel, produced by a very fast spark discharge," *Symp. Combust.*, vol. 22, no. 1, pp. 1651–1659, 1989.

[3] S. Stepanyan, N. Minesi, A. Tibere-Inglesse, A. Salmon, G. D. Stancu, and C. O. Laux, "Spatial evolution of the plasma kernel produced by nanosecond discharges in air," *J. Phys. D. Appl. Phys.*, vol. 52, no. 29, May 2019.

[4] M. Kono, K. Niu, T. Tsukamoto, and Y. Ujiie, "Mechanism of flame kernel formation produced by short duration sparks," in *Symposium (International) on Combustion*, 1989, vol. 22, no. 1, pp. 1643–1649.

[5] M. Thiele, J. Warnatz, & U. Maas, and U. Maas, "Geometrical study of spark ignition in two dimensions," *Combust. Theory Model.*, vol. 4, no. 4, pp. 413–434, Dec. 2000.

[6] F. Tholin and A. Bourdon, "Simulation of the hydrodynamic expansion following a nanosecond pulsed spark discharge in air at atmospheric pressure," *J. Phys. D-Applied Phys.*, vol. 46, no. 36, 2013.

[7] M. Akram, "Two-dimensional model for spark discharge simulation in air," *AIAA J.*, vol. 34, no. 9, pp. 1835–1842, Sep. 1996.

[8] M. Castela, B. Fiorina, A. Coussement, O. Gicquel, N. Darabiha, and C. O. Laux, "Modelling the impact of non-equilibrium discharges on reactive mixtures for simulations of plasma-assisted ignition in turbulent flows," *Combust. Flame*, vol. 166, pp. 133–147, 2016.

[9] B. Singh, L. K. Rajendran, P. P. Vlachos, and S. P. M. Bane, "Two regime cooling in flow induced by a spark discharge," *Phys. Rev. Fluids*, vol. 5, p. 14501, 2019.

[10] B. Singh, L. K. Rajendran, P. Gupta, C. Scalo, P. P. Vlachos, and S. P. Bane, "Experimental and Numerical Study of Flow Induced by Nanosecond Repetitively Pulsed Discharges," in *AIAA SciTech Forum*, 2019, no. January, pp. 1–15.

[11] B. Singh, L. K. Rajendran, M. Giarra, P. P. Vlachos, and S. P. M. Bane, "Measurement of the flow field induced by a spark plasma using particle image velocimetry," *Exp. Fluids*, vol. 59, no. 12, Dec. 2018.

[12] P. G. Saffman, "The Velocity of Viscous Vortex Rings," *Stud. Appl. Math.*, vol. 49, no. 4,

pp. 371–380, Dec. 1970.

[13] P. G. Saffman, *Vortex dynamics.* Cambridge university press, 1992.

[14] L. E. Fraenkel, "On steady vortex rings of small cross-section in an ideal fluid," 1970.

[15] J. Norbury, "A family of steady vortex," *J. Fluid Mech.*, vol. 57, pp. 417–431, 1973.

[16] H. Lamb, *Hydrodynamics*. Cambridge university press, 1895.

[17] C. Tung and L. Ting, "Motion and decay of a vortex ring," *Phys. Fluids*, vol. 10, no. 5, pp. 901–910, 1967.

[18] S. S. Yoon and S. D. Heister, "Analytical formulas for the velocity field induced by an infinitely thin vortex ring," *Int. J. Numer. Methods Fluids*, vol. 44, no. 6, pp. 665–672, 2004.

[19] H. Helmholtz, "On Integrals of the hydrodynamical equations, which express vortex-motion," *London, Edinburgh, Dublin Philos. Mag. J. Sci.*, vol. 33, pp. 485–512, 1867.

[20] M. N. Slipchenko, J. D. Miller, S. Roy, T. R. Meyer, J. G. Mance, and J. R. Gord, "100 kHz, 100 ms, 400 J burst-mode laser with dual-wavelength diode-pumped amplifiers.," *Opt. Lett.*, vol. 39, no. 16, pp. 4735–8, 2014.

[21] "No Title." [Online]. Available: https://github.com/aether-lab/prana.

[22] A. C. Eckstein and P. P. Vlachos, "Assessment of advanced windowing techniques for digital particle image velocimetry (DPIV).," *Meas. Sci. Technol.*, vol. 20, no. 7, p. 075402, Jul. 2009.

[23] C. Willert, "Stereoscopic digital particle image velocimetry for application in wind tunnel flows," *Meas. Sci. Technol.*, vol. 8, no. 12, pp. 1465–1479, 1999.

[24] M. C. Brindise and P. P. Vlachos, "Proper orthogonal decomposition truncation method for data denoising and order reduction," *Exp. Fluids*, vol. 58, no. 4, pp. 1–18, 2017.

[25] S. Bhattacharya, J. J. Charonko, and P. P. Vlachos, "Stereo-particle image velocimetry uncertainty quantification," *Meas. Sci. Technol.*, vol. 28, no. 1, 2017.

[26] S. Bhattacharya, J. J. Charonko, and P. P. Vlachos, "Particle image velocimetry (PIV) uncertainty quantification using moment of correlation (MC) plane," *Meas. Sci. Technol.*, vol. 29, no. 11, p. 115301, Nov. 2018.

[27] L. Rajendran, J. Zhang, S. Bane, and P. Vlachos, "Weighted Least Squares (WLS) Density Integration for Background Oriented Schlieren (BOS)," *arXiv Prepr. arXiv2004.01217*, pp. 1–14, Apr. 2020.

[28] L. K. Rajendran, J. Zhang, S. Bhattacharya, S. P. M. Bane, and P. P. Vlachos, "Uncertainty quantification in density estimation from background-oriented Schlieren measurements," *Meas. Sci. Technol.*, vol. 31, no. 5, Sep. 2020.

[29] A. Etebari and P. Vlachos, "Improvements on the accuracy of derivative estimation from DPIV velocity measurements," *Exp. Fluids*, pp. 39(6), 1040–1050, 2005.

[30] J. Zhou, R. J. Adrian, S. Balachandar, and T. M. Kendall, "Mechanisms for generating coherent packets of hairpin vortices in channel flow," *J. Fluid Mech.*, vol. 387, pp. 353-396,387, 1999.

[31] J. M. Wang, D. A. Buchta, and J. B. Freund, "Hydrodynamic ejection caused by laser-induced optical breakdown," *J. Fluid Mech.*, vol. 888, 2020.

[32] L. K. Rajendran, B. Singh, R. Jagannath, G. N. Schmidt, P. P. Vlachos, and S. P. Bane, "Experimental Characterization of Flow Induced by a Nanosecond Surface Discharge," 2020.

[33] A. Santhanakrishnan and J. D. Jacob, "Flow control with plasma synthetic jet actuators," *J. Phys. D. Appl. Phys.*, vol. 40, no. 3, pp. 637–651, 2007.